# Relieving the frustration through Mn$^{3+}$ substitution in Holmium Gallium Garnet

Paromita Mukherjee[1*], Hugh F.J. Glass[1], Emmanuelle Suard[2], Siân E. Dutton[1*]

[1] Cavendish Laboratory, University of Cambridge, JJ Thomson Avenue, Cambridge CB3 0HE, United Kingdom
[2] Institut Laue-Langevin, 71 Avenue des Martyrs, 38000 Grenoble, France

*corresponding authors: pm545@cam.ac.uk, sed33@cam.ac.uk

**Abstract:**

We present a study on the impact of Mn$^{3+}$ substitution in the geometrically frustrated Ising garnet Ho$_3$Ga$_5$O$_{12}$ using bulk magnetic measurements and low temperature powder neutron diffraction. We find that the transition temperature, $T_N$ = 5.8 K, for Ho$_3$MnGa$_4$O$_{12}$ is raised by almost 20 when compared to Ho$_3$Ga$_5$O$_{12}$. Powder neutron diffraction on Ho$_3$Mn$_x$Ga$_{5-x}$O$_{12}$ ($x$ = 0.5, 1) below $T_N$ shows the formation of a long range ordered ordered state with **k** = (0,0,0). Ho$^{3+}$ spins are aligned antiferromagnetically along the six crystallographic axes with no resultant moment while the Mn$^{3+}$ spins are oriented along the body diagonals, such that there is a net moment along [111]. The magnetic structure can be visualised as ten-membered rings of corner-sharing triangles of Ho$^{3+}$ spins with the Mn$^{3+}$ spins ferromagnetically coupled to each individual Ho$^{3+}$ spin in the triangle. Substitution of Mn$^{3+}$ completely relieves the magnetic frustration with $f = \theta_{CW}/T_N \sim 1.1$ for Ho$_3$MnGa$_4$O$_{12}$.

**Main text:**

In geometrically frustrated magnets (GFMs) the lattice geometry prevents all the magnetic interactions from being satisfied simultaneously. Two consequences of this are a large degeneracy in the number of possible ground states and a suppression of the long-range magnetic ordering temperature. Experimentally it has been observed that factors including symmetric and antisymmetric exchange, dipolar interactions, crystal electric field (CEF) effects, and lattice distortions play a role in determining the magnetic properties. Depending on the relative magnitude of competing interactions, the system may be driven into a long range ordered state, thus relieving the frustration, or exist in a disordered but correlated state such as spin liquid, or spin ice, or one with emergent magnetic order.[1–8] Magnetic frustration can also be relieved through site dilution or site disorder of spins.[9–13]

Lanthanide garnets with the general formula $Ln_3A_2X_3O_{12}$ are a system containing a highly frustrated magnetic $Ln^{3+}$ lattice. They crystallise in a cubic structure, Figure 1a, containing three crystallographic sites for the cations: dodecahedral occupied by $Ln$, octahedral occupied by $A$ and tetrahedral occupied by $X$. The magnetic $Ln^{3+}$ ions lie at the vertices of corner-sharing triangles which form two interpenetrating networks of bifurcated ten-membered rings, Figure 1b. The magnetic properties of the lanthanide garnets are highly dependent on the single ion anisotropy of the $Ln^{3+}$ ion and the cations on the octahedral and tetrahedral sites.[14–18] Much of the experimental and theoretical work so far has focused on the spin liquid candidate gadolinium gallium garnet (GGG), Gd$_3$Ga$_5$O$_{12}$.[15,19–22] Here we focus on the isostrucural holmium gallium garnet, Ho$_3$Ga$_5$O$_{12}$ (HoGG), which exhibits substantial single ion anisotropy.[23] Ho$_3$Ga$_5$O$_{12}$ was reported to undergo long-range magnetic ordering below 0.19 K in a six sublattice antiferromagnetic structure; however a later neutron scattering study points to coexistence of long and short-range magnetic order below 0.3 K down to 0.05 K.[24–26] We explore the impact of magnetic Mn$^{3+}$ substitution on the magnetic properties and magnetic structure of holmium gallium garnet.

We show that the magnetic frustration of the Ising garnet Ho$_3$Ga$_5$O$_{12}$ is almost entirely relieved by partial substitution of nonmagnetic Ga$^{3+}$ with magnetic Mn$^{3+}$. In the case of Ho$_3$Mn$_x$Ga$_{5-x}$O$_{12}$ ($x$ = 0.5, 1), the Mn$^{3+}$ spins create a local dipolar field, coupling ferromagnetically with *quasi-spins* from Ho$_3$ triangles. The Mn$^{3+}$ spins and the Ho$_3$ *quasi-spin* sublattices in Ho$_3$MnGa$_4$O$_{12}$ form a long range ordered state at $T_N$ = 5.8 K, a dramatic contrast to the reported coexistence of short and long range order observed below 0.3 K for unsubstituted Ho$_3$Ga$_5$O$_{12}$.[26]

Polycrystalline samples of phase pure Ho$_3$Mn$_x$Ga$_{5-x}$O$_{12}$ ($0 \leq x \leq 1$) have been prepared and the structure evaluated using X-ray and neutron diffraction as described in the supplementary material. Mn$^{3+}$ substitution results in a small increase in the unit cell, however no significant changes in the Ho-O bond lengths are observed, Table S1 and Table S2. Analysis of the crystal structure shows that Mn$^{3+}$ exclusively occupies the octahedral *A* sites, located above and below each Ho$_3$ triangle (Figure 1c). The preference of $d^4$ Mn$^{3+}$ to occupy only the octahedral sites is expected from consideration of the CEF for the octahedral *A* and tetrahedral *X* sites. No evidence for ordering of the Mn$^{3+}$ ions or a Jahn-Teller distortion is observed, although local Jahn-Teller distortions cannot be discounted. At the maximum substitution, 50% of the *A* sites are occupied by magnetic Mn$^{3+}$ ions. The connectivity of the *A* sites has been described by one-dimensional chains propagating along the body diagonal of the cubic unit cell,[27] however all the sites occupied by Mn$^{3+}$ spins, including those in neighbouring chains, are equidistant from one another in the unit cell.

The Zero Field Cooled (ZFC) magnetic susceptibility, $\chi(T)$, of Ho$_3$Mn$_x$Ga$_{5-x}$O$_{12}$ ($0 \leq x \leq 1$), Figure 2a, shows a sharp magnetic ordering transition, $T_N$, at 3.5 K and 5.8 K for Ho$_3$Mn$_{0.5}$Ga$_{4.5}$O$_{12}$ and Ho$_3$MnGa$_4$O$_{12}$ respectively. No ordering is observed in Ho$_3$Ga$_5$O$_{12}$ above the limiting temperature of 1.8 K, consistent with previous literature reports.[26,28] The inverse susceptibility, $\chi^{-1}$, is linear at high temperatures, $T > 100$ K (Figure 2a inset) and fits to the Curie-Weiss law were carried out in different temperature ranges from 100 - 300 K. The difficulty in determining the Weiss temperature, $\theta_{CW}$, from high temperature fits to the Curie-Weiss law is well documented for Ho$^{3+}$ containing samples due to the presence of low-lying crystal electric field (CEF) states.[23,26,29] However, for all compositions, $\theta_{CW}$ is negative, indicating net antiferromagnetic interactions. The value of $\theta_{CW}$ decreases with increase in $x$, indicating weaker antiferromagnetic correlations on Mn$^{3+}$ substitution. The effective moment, $\mu_{eff}$, obtained from the Curie Weiss law, Table S3, is underestimated compared to the theoretical moment [a]($\mu_{th}^2 = 3\mu_{Ho}^2 + x\mu_{Mn}^2$). However, $\mu_{eff}$ increases with $x$, as expected for Mn$^{3+}$ substitution.

Isothermal magnetisation curves (Figure 2b), show that the magnetisation at 2 K and 9 T, $M_{2K,9T}$, is significantly increased on Mn$^{3+}$ substitution. The size of the increase cannot solely be attributed to the Mn$^{3+}$ ions as it exceeds the maximum contribution from Mn$^{3+}$ ($M_{\text{Mn max}} = g_S S = 4$ μ$_B$ per formula unit). The additional increase in magnetisation could be due to changes in the underlying magnetism or in the CEF states of Ho$^{3+}$ on substitution. For all samples the observed magnetisation at 9 T, $M_{2K,9T}$, is much lower than the saturation magnetisation of a Heisenberg system, $M_{sat} = 3 \times 10 + x \times 4$ μ$_B$/f.u. ($3g_J J + x g_S S$ where $g_J = 5/4$, $J = 8$ for Ho$^{3+}$ and $g_S = 2$, $S = 2$ for Mn$^{3+}$). However, it is consistent with the value expected for powder averaged Ising Ho$^{3+}$ spins; $M_{sat,Ising} = 3 \times 10/2 + x \times 4$ μ$_B$/f.u. The isothermal magnetisation in Ho$_3$Ga$_5$O$_{12}$ has previously been shown to be typical of Ising spins,[23] and our data is consistent with the Ho$^{3+}$ spins remaining Ising-like on Mn substitution. Given their small contribution to the total magnetisation, no conclusions can be drawn regarding the isotropy of the Mn$^{3+}$ spins. At 2 K, a field-induced transition is observed at 0.27(1) T and 0.46(1) T for Ho$_3$Mn$_{0.5}$Ga$_{4.5}$O$_{12}$ and Ho$_3$MnGa$_4$O$_{12}$ respectively, Figure S3. Similar transitions in Ising garnets containing magnetic ions exclusively on the *A* site have recently been reported.[27] The plot of *dM/dH* for Ho$_3$Ga$_5$O$_{12}$ also shows a feature at low fields < 0.2 T, Figure S3, however further measurements are required to understand the nature of these field-induced transitions.

---

[a] Assumes no quenching of the orbital contribution to the effective moment. However, partial quenching of the moment would be expected due to presence of low-lying CEF states.

To explore the nature of the magnetic ordering, we carried out low temperature powder neutron diffraction (PND) experiments on Ho$_3$Mn$_{0.5}$Ga$_{4.5}$O$_{12}$ and Ho$_3$MnGa$_4$O$_{12}$. Both samples show strong magnetic Bragg peaks below $T_N$. No magnetic diffuse scattering is observed for either sample at $T \geq 1.5$ K suggesting that unlike in HoGG,[26] long and short-range magnetic order do not coexist. For both samples, the magnetic Bragg reflections are indexed with the propagation vector **k** = (0, 0, 0). All combinations of irreducible representations for Ho$^{3+}$ and Mn$^{3+}$ ions were tested, however only a model with both ions having the $\Gamma_3^1$ irreducible representation, Table S4, allowed for a good fit to the data (Figure 3a). For both samples, the magnitude of the Ho$^{3+}$ and Mn$^{3+}$ moments increase on cooling, though the moments are smaller than the theoretical moment$^a$ ( $g_J\sqrt{J(J+1)} = 10.61 \mu_B$ for Ho$^{3+}$ and $g_S\sqrt{S(S+1)} = 4.89$ $\mu_B$ for Mn$^{3+}$ respectively), Figure S4. This may be due to low-lying CEF effects or screening of the moment. Previous studies of Ho$_3$Ga$_5$O$_{12}$ and Ho$_3$Al$_5$O$_{12}$ have also reported reduced moments, in close agreement with our results.[28,30] Reduced magnetic moments for Mn$^{3+}$ determined from neutron diffraction have also previously been observed.[31]

The magnetic structure, Figure 3b, has the same long range ordered arrangement of the Ho$^{3+}$ spins as that reported for Ho$_3$Ga$_5$O$_{12}$ and Ho$_3$Al$_5$O$_{12}$.[25,30] The 24 Ho$^{3+}$ spins in each unit cell are arranged into six sublattices with the Ho$^{3+}$ spins aligned along the crystallographic axes [100], [$\bar{1}$00], [010], [0$\bar{1}$0], [001] and [00$\bar{1}$] such that the net moment is zero. The Mn$^{3+}$ spins in each unit cell are aligned along the body diagonals, as reported for the Ising garnet CaY$_2$Co$_2$Ge$_3$O$_{12}$,[27] however their relative orientation is completely different. The Mn$^{3+}$ are oriented along [111], [$\bar{1}$1$\bar{1}$], [$\bar{1}$11$\bar{1}$] and [1$\bar{1}\bar{1}$] such that there is a resultant moment of from the Mn$^{3+}$ spins along [111]. The relative orientations of the Ho$^{3+}$ and Mn$^{3+}$ spins assume greater significance when we consider the two interpenetrating networks of ten-membered triangles of Ho$^{3+}$ spins, Figure 3c. For each ten-membered ring, the net magnetic moment of the Ho$^{3+}$ spins is zero, however there is a net ferromagnetic interaction between the Ho$^{3+}$ and Mn$^{3+}$ moments. When these interactions are summed over a Ho$_3$ triangle then the resultant Ho$^{3+}$ *quasi-spin* is orientated in or out of the centroid of the triangle, i.e. along [111] (Figure 3d) and are located directly above or below the site partially occupied by $Mn^{3+}$. The $Mn^{3+}$ spin aligns co-parallel with the $Ho_3$ *quasi-spin* (Figure 3e). Whilst the construct of the $Ho_3$ *quasi-spins* allows for the magnetic structure to be rationalised it should be noted that in the parent material, $Ho_3Ga_5O_{12}$, coupling between any two of the $Ho^{3+}$ spins on an individual triangle $\propto \boldsymbol{S_1}.\boldsymbol{S_2}$ results in no net interaction as they are orthogonal, however in the case of $Ho_3Mn_xGa_{5-x}O_{12}$ each individual $Ho^{3+}$-$Mn^{3+}$ interaction is non-zero.

To our knowledge the concurrent magnetic ordering observed of both $Ho^{3+}$ and $Mn^{3+}$ in $Ho_3MnGa_4O_{12}$ is unique when compared to other rare-earth–transition metal oxides with complex magnetic structures. Studies on magnetic dopants in lanthanide garnets have been restricted to $Ln_3Fe_5O_{12}$ where $Fe^{3+}$ occupies both octahedral and tetrahedral sites. The two $Fe^{3+}$ sublattices order in a ferrimagnetic structure at ~ 130 – 140 K while the $Ln^{3+}$ ions order in an umbellate structure around ~ 10 K.[32–34] In $HoMnO_3$, the $Mn^{3+}$ spins order at ~ 72 K while the onset of ordering in the $Ho^{3+}$ spins is seen at the spin-rotation transition for the $Mn^{3+}$ spins ~33 K followed by an increase in the ordered $Ho^{3+}$ moments below 5 K.[35–37] However, in $Ho_3MnGa_4O_{12}$, no features are observed in the magnetic susceptibility or neutron diffraction data corresponding to individual ordering of the $Mn^{3+}$ spins at $T > T_N$. The ordering mechanism is also distinct from the 'ordered spin-ice' structure reported for $Ho_2CrSbO_7$, where the frustration is proposed to be relieved through local ferromagnetic correlations between the $Cr^{3+}$ spins, as is evidenced by a positive Curie-Weiss constant for isostructural $Y_2CrSbO_7$.[13,38] However, in $Ho_3MnGa_4O_{12}$, the Mn-Mn and Ho-Ho exchange interactions are antiferromagnetic suggesting that the ordering is driven by a different mechanism, the origin of which is discussed below.

The partial substitution of $Ga^{3+}$ for $Mn^{3+}$ in $Ho_3Mn_xGa_{5-x}O_{12}$ significantly changes the magnetic interactions which need to be considered. In addition to Ho-Ho interactions present in $Ho_3Ga_5O_{12}$, Mn-Mn and Ho-Mn interactions also need to be considered. First we consider the dipolar and exchange interactions between the magnetic $Ho^{3+}$. As the Ho-Ho bond lengths are not significantly changed on $Mn^{3+}$ substitution, it can be assumed that there is no significant change in the Ho-Ho dipolar interaction energy, $D \sim \frac{\mu_0 \mu_{eff}^2}{4\pi R_{nn}^3} \sim$ 0.9 K. A priori calculation of the Ho-Ho exchange interactions is complex, as the Curie-Weiss constants for the $Mn^{3+}$ substituted garnets contain contributions from multiple interactions. An order of magnitude approximation for the nearest-neighbour exchange energy, $J_1$, in unsubstituted $Ho_3Ga_5O_{12}$ can be obtained as $J_1 \sim \frac{3k_B\theta_{CW}}{2n}$ where $n$ = number of nearest-neighbour $Ho^{3+}$ = 4. This gives $J_1$ = - 4.5 K and an order of magnitude estimation of $J_1$ for $Ho_3Mn_xGa_{5-x}O_{12}$. The Mn-Mn exchange interactions can be approximated by considering isostructural $Y_3MnGa_4O_{12}$ (with analogous lattice parameter and bond lengths as $Ho_3MnGa_4O_{12}$, Table S1 and Table S2); here the only magnetic contribution is from the $Mn^{3+}$ spins. The magnetic susceptibility of $Y_3MnGa_4O_{12}$ is shown in Figure 2c. The divergence in the Zero Field-Cooled and Field-Cooled data at $T_f$ = 18 K is characteristic of spin-glass like behaviour. Given the site disorder, formation of a spin-glass state is not unexpected and has been observed in other systems with dilute spins along [111].[39] Fits to the Curie-Weiss law between 100 – 300 K gives $\mu_{eff}$ = 4.83 $\mu_B$ consistent with $Mn^{3+}$ spins and $\theta_{CW}$ = -9(4) K, indicating antiferromagnetic interactions between $Mn^{3+}$. This corresponds to $J_1 \sim$ - 6.8 K if each $Mn^{3+}$ spin is assumed to have 2 nearest neighbours. Determination of the Ho-Mn exchange interactions is nontrivial and further inelastic neutron scattering experiments are required for quantitative analysis.

However, the resultant spin structure, although constrained by CEF effects, has a ferromagnetic component between adjacent $Ho^{3+}$ and $Mn^{3+}$ spins, suggesting the resulting moment is not minimised. Finally we consider the Ho-Mn dipolar interactions. The local internal dipolar fields due to the $Mn^{3+}$ spins above and below the $Ho_3$ triangles can be approximated as $\mu_0 H \sim \frac{\mu_0 \mu_{eff}}{2\pi r^3} = \frac{\mu_0 g_S \sqrt{S(S+1)} \mu_B}{2\pi r^3}$ where $g_S = 2$, $S = 2$ for $Mn^{3+}$ and $r$ is the distance between the centroid of the $Ho_3$ triangle and the $Mn^{3+}$ spin = 2.65 Å ~ 0.5 T, this corresponds to an energy ~ 3.2 K. We find a direct relationship between $T_N$ and the number of *quasi-spins* experiencing a local magnetic field (Figure 3a inset) when a random distribution of $Mn^{3+}$ is assumed. This indicates that the local internal dipolar field may play a role in the magnetic ordering. In $Ho_3Ga_5O_{12}$, the formation of a long range ordered state is observed on application of a 2 T field along [111],[26] this can be interpreted as equivalent to 25% of the $Ho_3$ triangles experiencing a local field. Whilst the nature of field-induced long-range ordering in $Ho_3Ga_5O_{12}$ is unknown, this highlights the role of an applied field in the magnetic ordering in Ising garnets.

In conclusion we find that in $Ho_3Mn_{0.5}Ga_{4.5}O_{12}$ and $Ho_3MnGa_4O_{12}$, the $Mn^{3+}$ moments, disordered on the octahedral site, couple ferromagnetically with the $Ho_3$ *quasi-spins* and lift the degeneracy associated with magnetic ordering in Ising garnets. The elevation of the ordering temperature almost completely relieves the magnetic frustration, $f = |\theta_{CW}/T_N|$[1] such that $f \sim 1.1$ for $Ho_3MnGa_4O_{12}$ compared to $f \sim 40$ for $Ho_3Ga_5O_{12}$, Table S3. Susceptibility measurements show similar increases in $T_N$ for $Ln_3Mn_xGa_{5-x}O_{12}$ ($Ln$ = Tb, Dy). The $Cr^{3+}$ substituted lanthanide gallium garnets, $Ln_3CrGa_4O_{12}$ ($Ln$ = Tb, Dy, Ho), also show an increase in $T_N$ by a smaller factor than on $Mn^{3+}$ substitution.[40] Neutron diffraction is required to elucidate the magnetic structure in these cases but this hints at a universal mechanism for relieving the magnetic frustration in Ising lanthanide garnets which is tuneable through control of the extent and type of magnetic ion substitution.

**Acknowledgements**

We thank J. Hodkinson for his support during the experiments on D1B, ILL. We acknowledge funding support from the Winton Programme for the Physics of Sustainability. Magnetic measurements were carried out using the Advanced Materials Characterisation Suite, funded by EPSRC Strategic Equipment Grant EP1M00052411.

Supporting data can be found at https://doi.org/10.17863/CAM.13758, neutron diffraction data can also be found at ILL https://doi.org/10.5291/ILL-DATA.5-31-2457.

**Figure Captions**

**Figure 1.** (Colour online) a) General crystal structure of lanthanide garnets $Ln_3A_2X_3O_{12}$ with the three cations occupying distinct crystallographic sites– here $Ln$ = Ho, $A$ = Mn/Ga, $X$ = Ga b) Connectivity of magnetic $Ho^{3+}$ ions. The $Ho^{3+}$ lie at the vertices of corner-sharing equilateral triangles forming two interpenetrating ten-membered rings. This results in a highly frustrated three-dimensional network c) Relative position of $Mn^{3+}$ relative to $Ho^{3+}$ - each triangle with $Ho^{3+}$ at the vertices has a $Mn^{3+}$ atom above and below the centroid of the triangle. Each octahedral site is occupied by $Mn^{3+}$ 25% and 50% of the time for $Ho_3Mn_{0.5}Ga_{4.5}O_{12}$ and $Ho_3MnGa_4O_{12}$ respectively

**Figure 2.** (Colour online) a) Zero field cooled (ZFC) magnetic susceptibility $\chi(T)$ measured in 100 Oe for $Ho_3Mn_xGa_{5-x}O_{12}$ ($0 \leq x \leq 1$): Magnetic ordering transitions are clearly seen at 3.5 K and 5.8 K for $x = 0.5$ and $x = 1$ respectively. The inverse magnetic susceptibility, $\chi^{-1}$, is inset. b) Isothermal magnetisation curves at 2 K for $Ho_3Mn_xGa_{5-x}O_{12}$ ($0 \leq x \leq 1$) c) ZFC and field cooled (FC) magnetic susceptibility $\chi(T)$ measured in 100 Oe for $Y_3MnGa_4O_{12}$: A broad spin-glass-like transition is observed at $T_0 = 18$ K. The inverse magnetic susceptibility, $\chi^{-1}(T)$, is inset.

**Figure 3.** (Colour online) a) Rietveld refinement of the neutron diffraction pattern at 1.5 K for $Ho_3MnGa_4O_{12}$: Blue ticks – nuclear Bragg reflections, red ticks – magnetic Bragg reflections; inset shows the ordering temperature, $T_N$, as a function of % of $Ho_3$ triangles experiencing the local internal field from the $Mn^{3+}$ spins. b) Magnetic structure for $Ho_3MnGa_4O_{12}$ ($T_N = 5.8$ K) c) Arrangement of $Ho^{3+}$ and $Mn^{3+}$ spins for $Ho_3MnGa_4O_{12}$ in the two interpenetrating ten-membered rings in the garnet lattice d) Each $Ho_3$ triangle has three orthogonal spins orientated along the three crystallographic axes, the $Ho^{3+}$ *quasi-spin* directed along [111] is also shown. e) The $Ho^{3+}$ *quasi-spin* couples ferromagnetically with the $Mn^{3+}$ spins located above and below the triangle.

**Figures**

Figure 1

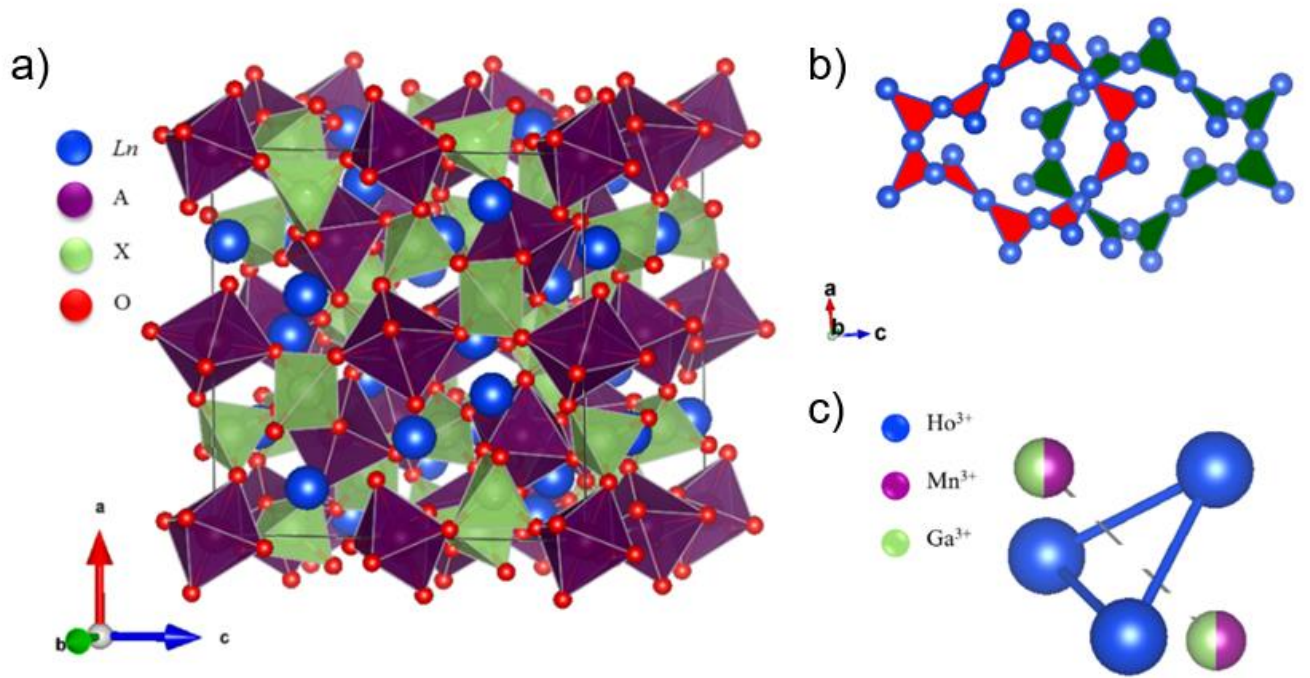

Figure 2

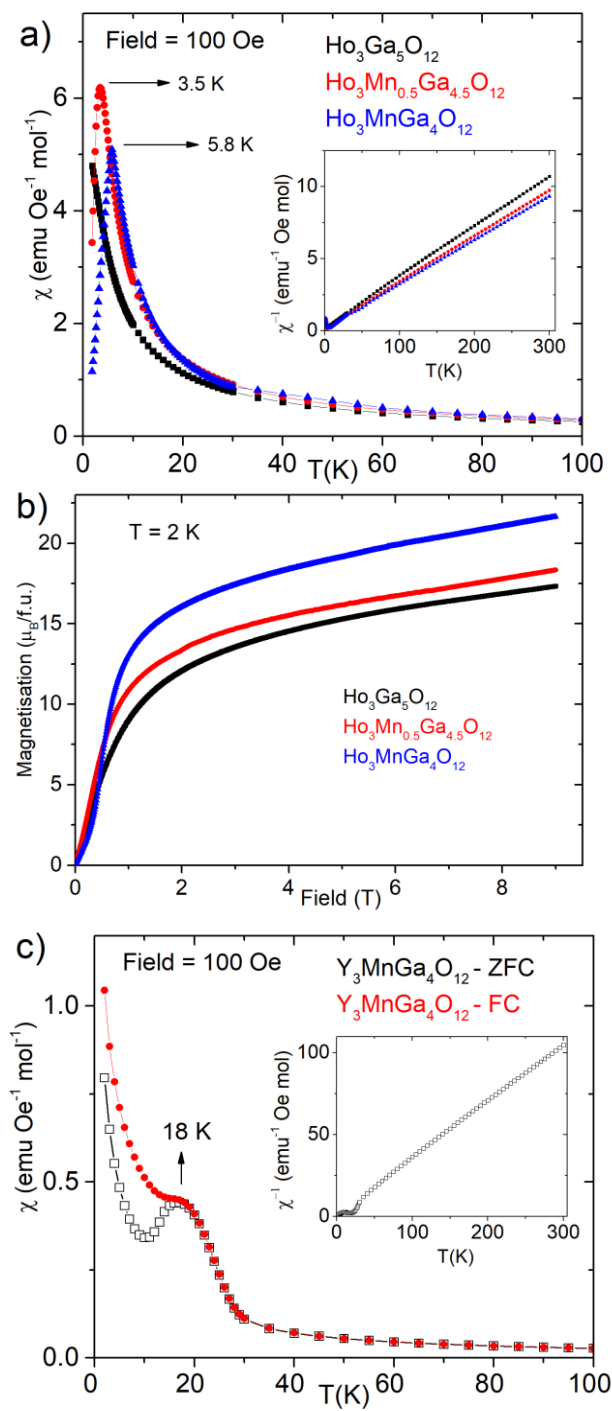

**Supplementary information**

**1. Sample preparation and experimental methods:**

Powder samples of $Ho_3Mn_xGa_{5-x}O_{12}$ ($x = 0, 0.5, 1$) and $Y_3MnGa_4O_{12}$ were prepared using a solid-state synthesis route. Samples were made by mixing stoichiometric amounts of $Ho_2O_3$ (99.999%, Alfa Aesar) or $Y_2O_3$ (99.999% Alfa Aesar), $Ga_2O_3$ (99.999%, Alfa Aesar), and $MnO_2$ (99.999%, Alfa Aesar). To ensure the correct stoichiometry $Ga_2O_3$ was pre-dried at 500 °C. Pellets were heated in air at 1200 °C repeatedly for 48-72 hours with intermittent regrindings. A reaction was deemed completed when powder X-Ray diffraction (PXRD) indicated the formation of a phase pure product.

Structural analysis was carried out using PXRD. Initially short scans were collected over $10° \leq 2\theta \leq 60°$ ($\Delta 2\theta = 0.015°$) using a Bruker D8 X-Ray diffractometer (Cu Kα radiation, $\lambda = 1.540$ Å). For quantitative structural analysis, longer scans for 2 hours over a wide angular range $10° \leq 2\theta \leq 90°$ ($\Delta 2\theta = 0.01°$) were collected. For $Ho_3MnGa_4O_{12}$, room temperature PND experiments for structural characterisation were carried out on the D2B diffractometer, ILL ($\lambda = 1.595$ Å). PXRD and PND structural Rietveld refinements were carried out using the Fullprof suite of programmes.[41] Backgrounds were fitted using linear interpolation and the peak shape was modelled using a pseudo-Voigt function.

Magnetic measurements were carried out on a Quantum Design Magnetic Properties Measurement System (MPMS) with a Superconducting Quantum Interference Device (SQUID). The zero-field cooled (ZFC) magnetic susceptibility, $\chi(T)$, was measured in a field of 100 Oe in the temperature range 1.8 - 300 K for $Ho_3Mn_xGa_{5-x}O_{12}$ ($0 \leq x \leq 1$) and $Y_3MnGa_4O_{12}$. In a low field of 100 Oe, the isothermal magnetisation $M(H)$ curve is linear at all $T$ and so the linear approximation for $\chi(T)$ is valid i.e. $\chi(T) \sim M/H$. $M(H)$ measurements in the field range, $\mu_oH = 0 – 9$ T for selected temperatures were made on all samples using the ACMS (AC Measurement System) option on a Quantum Design Physical Properties Measurement System (PPMS).

In order to solve the long range ordered magnetic structure of $Ho_3Mn_xGa_{5-x}O_{12}$ ($x = 0.5, 1$), low temperature PND measurements, $T \geq 1.5$ K, were carried out on the D1B ($\lambda = 2.525$ Å) and D20 ($\lambda = 1.542$ Å) diffractometers at ILL, Grenoble. The magnetic cell was indexed using the k-search program in the Fullprof suite. Different combinations of irreducible representations for $Ho^{3+}$ and $Mn^{3+}$ were tested using the SARAH program.[42] The irreducible representations were combined using SARAH to generate a single magnetic phase for the magnetic Rietveld refinement in Fullprof.

## 2. Structural Rietveld refinement of $Ho_3Mn_xGa_{5-x}O_{12}$ ($x$ = 0.5, 1):

Quantitative analysis of the crystal structure for $Ho_3MnGa_4O_{12}$ was carried out using a combined Rietveld analysis of room temperature PXRD and PND patterns. For $Ho_3Ga_5O_{12}$ and $Ho_3Mn_{0.5}Ga_{4.5}O_{12}$, the crystal structure parameters were determined from room temperature PXRD alone; however for the latter, the Mn occupancy was determined from the low temperature PND pattern at 15 K. On substitution with Mn, the lattice parameter is observed to increase slightly. Mn is expected to occupy the octahedral sites from crystal electric field considerations for $d^4$ $Mn^{3+}$ ions. PND is highly sensitive to the positions and amount of Mn and Ga in the structure due to the contrast between the neutron scattering lengths of manganese, $b_{Mn}$ = -3.73 fm, and gallium, $b_{Ga}$ = 7.29 fm. Structural models considering Mn on both octahedral and tetrahedral sites were considered, only the model with Mn exclusively on the octahedral site gave a good fit. The fractional occupancy of the octahedral site was refined for the Mn doped Ho garnets. The composition of the Mn doped garnets as determined from neutron diffraction was found to be $Ho_3Mn_{0.46(2)}Ga_{4.54(2)}O_{12}$ and $Ho_3Mn_{1.12(2)}Ga_{3.88(2)}O_{12}$ for nominal compositions of x = 0.5 and x = 1 respectively. The large neutron scattering length of oxygen, $b_O$ = 5.80 fm, also enabled the possibility of oxygen vacancies in the lattice to be explored, however no significant oxygen vacancies were observed and so the oxygen site was assumed to be fully occupied.

## 3. Structural characterisation:

Table S1 – Refined room temperature structural parameters for $Ho_3Mn_xGa_{5-x}O_{12}$ ($0 \leq x \leq 1$) and $Y_3MnGa_4O_{12}$. Ho/Y occupy the dodecahedral 24$c$ (0, 0.25, 0.125) site. Ga occupies the 24$d$ (0, 0.25, 0.375) tetrahedral site while Mn/Ga are disordered over the octahedral (0,0,0) 16$a$ site. The general ($x,y,z$) 96$h$ position is occupied by O.

* Parameters determined from PXRD+PND

**composition fixed from analysis of PND pattern at 15 K.

| $Ho_3Mn_xGa_{5-x}O_{12}$ | | \multicolumn{3}{c}{$x$} | | | $Y_3MnGa_4O_{12}$ |
|---|---|---|---|---|---|
| $Ia\bar{3}d$ | | 0 | 0.5 | 1.0* | |
| $a$ (Å) | | 12.28157(5) | 12.29232(11) | 12.3049(3) | 12.29337(4) |
| $\chi^2$ | | 4.32 | 3.70 | 2.55 | 5.56 |
| $R_{wp}$ | | 5.33 | 5.08 | 4.32 | 10.5 |
| Mn/Ga1 16$a$ (0, 0, 0) | Frac Mn | 0 | 0.23(2)** | 0.56(2) | 0.5 |
| O 96$h$ ($x,y,z$) | $x$ | -0.0298(2) | -0.02752(8) | -0.02796(7) | -0.0272(3) |
| | $y$ | 0.0515(3) | 0.05503(7) | 0.05579(8) | 0.0550(3) |
| | $z$ | 0.1494(3) | 0.150041(12) | 0.15044(8) | 0.1510(3) |

Table S2 – Selected bond lengths from room temperature powder X-ray diffraction refinements for $Ho_3Mn_xGa_{5-x}O_{12}$ ($0 \leq x \leq 1$). Bond lengths were determined from analysis of room temperature PXRD only to enable consistent comparisons.

| $Ho_3Mn_xGa_{5-x}O_{12}$ | $x$ | | | $Y_3MnGa_4O_{12}$ |
|---|---|---|---|---|
| $Ia\bar{3}d$ | 0 | 0.5 | 1.0 | |
| Ho-Ho (Å) | 3.76045(4) × 4 | 3.76374(3) × 4 | 3.76692(3) × 4 | 3.76406(4) × 4 |
| Ho-O (Å) | 2.354(4) × 4 | 2.378(4) × 4 | 2.346(4) × 4 | 2.332(4) × 4 |
| | 2.483(3) × 4 | 2.469(3) × 4 | 2.474(3) × 4 | 2.441(4) × 4 |
| <Ho-O> (Å) | 2.418 | 2.424 | 2.410 | 2.386 |
| Mn/Ga1-O (Å) | 1.975(3) × 6 | 1.960(3) × 6 | 1.987(3) × 6 | 2.004(4) × 6 |
| Ga2-O (Å) | 1.815(4) × 4 | 1.834(4) × 4 | 1.827(4) × 4 | 1.840(4) × 4 |

Figure S1 – Room temperature PXRD and PND pattern for Ho$_3$MnGa$_4$O$_{12}$: Experimental data (red dots), Modelled data (black line), Difference pattern (blue line), Bragg positions (blue ticks).

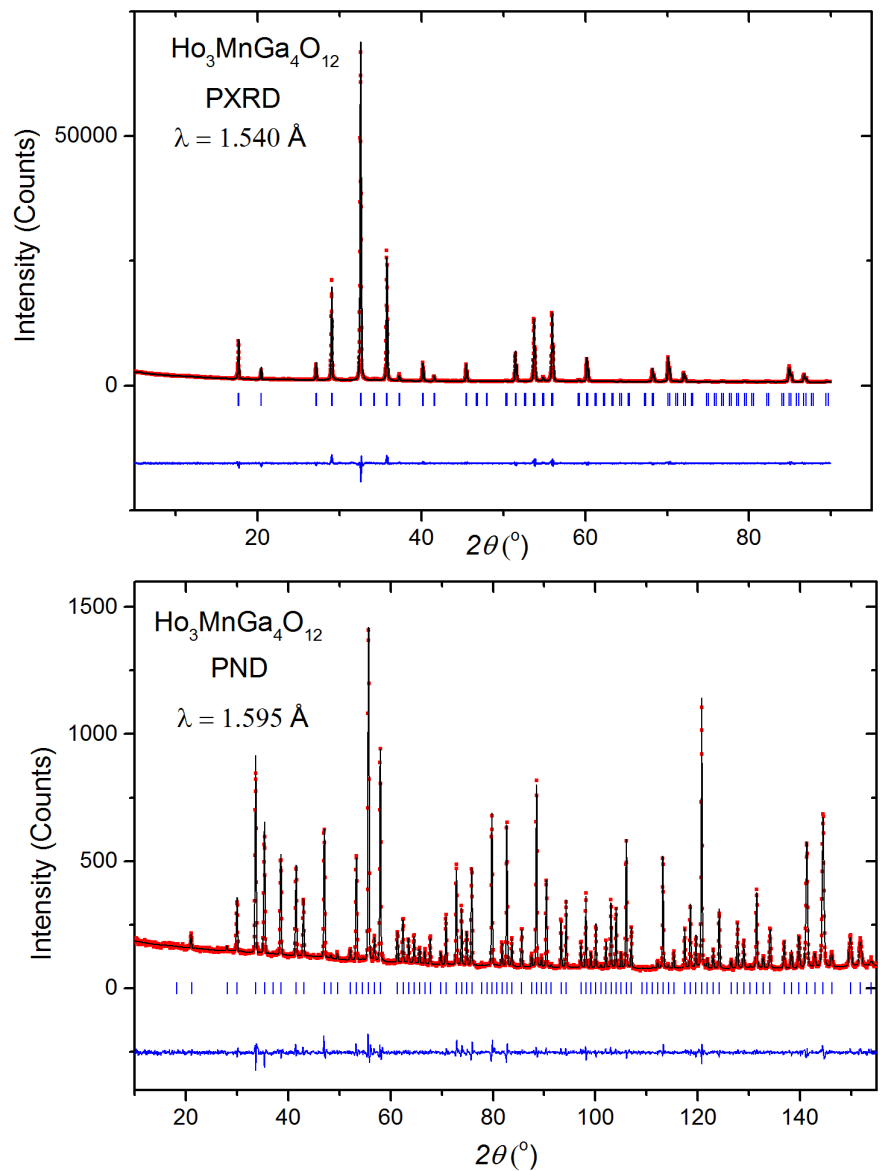

## 4. Bulk magnetic measurements:

Table S3 – Magnetisation parameters for $Ho_3Mn_xGa_{5-x}O_{12}$ ($0 \leq x \leq 1$). *$Ho_3Ga_5O_{12}$ has been reported to undergo an ordering transition below 0.3 K.[24,26]

| $Ho_3Mn_xGa_{5-x}O_{12}$ | | | | Theoretical | | Experimental | |
|---|---|---|---|---|---|---|---|
| $x$ | $T_N$ (K) | $\theta_{CW}$ (K) | $f = \left\|\dfrac{\theta_{CW}}{T_N}\right\|$ | $\mu_{eff}$ ($\mu_B$/f.u.) | $M_{sat} = g_J J$ ($\mu_B$/f.u.) | $\mu_{eff}$ ($\mu_B$/f.u.) | $M_{2K,9T}$ ($\mu_B$/f.u.) |
| 0 | < 1.8* | -12(4) | 40 | 18.4 | 30 | 15.2(2) | 17.2 |
| 0.5 | 3.5 (2) | -9(2) | 2.5 | 18.7 | 32 | 15.88(12) | 18.1 |
| 1.0 | 5.8 (2) | -6(3) | 1.1 | 19.0 | 34 | 16.1(2) | 21.1 |

Figure S2 – Isothermal magnetisation curves at selected temperatures for $Ho_3Mn_xGa_{5-x}O_{12}$ ($0 \leq x \leq 1$)

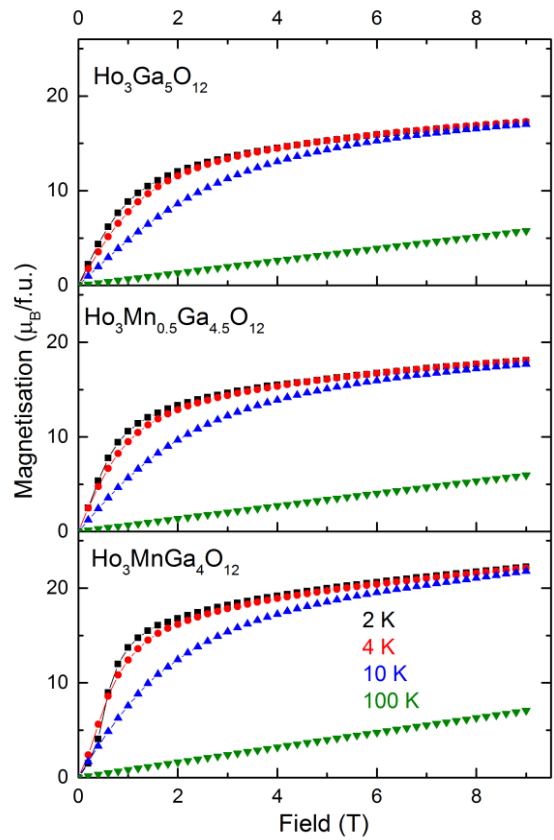

Figure S3 – Derivative of the magnetisation at $T = 2$ K for Ho$_3$Mn$_x$Ga$_{5-x}$O$_{12}$ ($0 \leq x \leq 1$) from 0 – 5 T, clear features are observed at 0.27(1) T and 0.46(1) T for $x = 0.5$ and $x = 1$ respectively

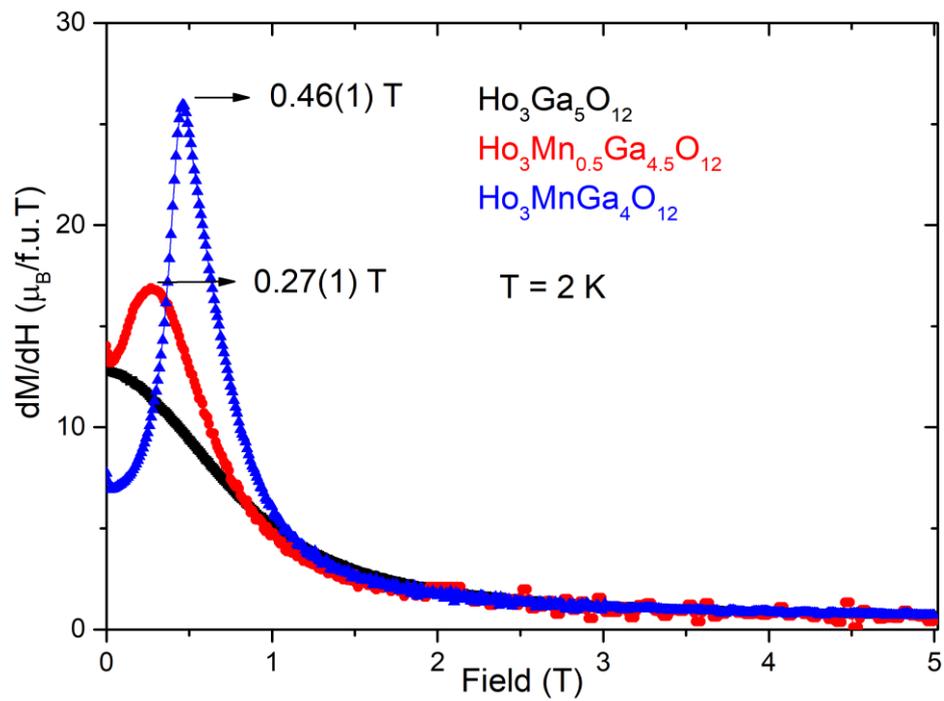

## 5. Magnetic Rietveld refinement of Ho$_3$Mn$_x$Ga$_{5-x}$O$_{12}$ ($x$ = 0.5, 1):

Symmetry analysis for the propagation vector **k** = (0, 0, 0) and space group $Ia\bar{3}d$ gave eight nonzero irreducible representations (IRs) for the magnetic Ho(24$c$) site: 2 one-dimensional ($\Gamma_3^1$, $\Gamma_4^1$) and 2 two-dimensional ($\Gamma_5^2$, $\Gamma_6^2$) all occurring once, 2 three-dimensional ($\Gamma_7^3$, $\Gamma_8^3$) occurring twice and 2 three-dimensional representations ($\Gamma_9^3$, $\Gamma_{10}^3$) that are repeated thrice. According to Kovalev's notation,[43] the magnetic representation $\Gamma_{\text{mag Ho}}$ is given by:

$$\Gamma_{\text{mag Ho}} = 1\Gamma_3^1 + 1\Gamma_4^1 + 1\Gamma_5^2 + 1\Gamma_6^2 + 2\Gamma_7^3 + 2\Gamma_8^3 + 3\Gamma_9^3 + 3\Gamma_{10}^3$$

Of these, only $\Gamma_5^2$ and $\Gamma_6^2$ have both real and imaginary components while the others only have real components. Similar representational analysis for the magnetic Mn(16$a$) site gave five nonzero IRs: 2 one-dimensional ($\Gamma_1^1$, $\Gamma_3^1$) repeated once, 1 two-dimensional ($\Gamma_6^2$) repeated twice and 2 three-dimensional IRs ($\Gamma_8^3$, $\Gamma_{10}^3$) repeated thrice in the decomposition. Then in the same notation $\Gamma_{\text{mag Mn}}$ is given by:

$$\Gamma_{\text{mag Mn}} = 1\Gamma_1^1 + 1\Gamma_3^1 + 2\Gamma_6^2 + 3\Gamma_8^3 + 3\Gamma_{10}^3$$

Of these, only $\Gamma_6^2$ has both real and imaginary components and all the others have real components only.

Different combinations of IRs were tried out and the $\Gamma_3^1$ representation for both Ho and Mn was found to give the best fit.

Table S4 – The basis vectors $\psi_1$ for Ho (24c) and Mn(16a) in nonzero IR $\Gamma_3$[1]

| Atoms in non-primitive basis for Ho | | Components of $\psi_1$ | Atoms in non-primitive basis for Mn | | Components of $\psi_1$ |
|---|---|---|---|---|---|
| Atom label | Coordinates | | Atom label | Coordinates | |
| Ho1 | (0, ¼, ⅛) | (0 0 1) | Mn1 | (0, 0, 0) | (1 1 1) |
| Ho2 | (0, ¾, ⅜) | (0 0 -1) | Mn2 | (½, 0, ½) | (-1 -1 1) |
| Ho3 | (⅛, 0, ¼) | (1 0 0) | Mn3 | (0, ½, ½) | (-1 1 -1) |
| Ho4 | (⅜, 0, ¾) | (-1 0 0) | Mn4 | (½, ½, 0) | (1 -1 -1) |
| Ho5 | (¼, ⅛, 0) | (0 1 0) | Mn5 | (¾, ¼, ¼) | (-1 -1 1) |
| Ho6 | (¾, ⅜, 0) | (0 -1 0) | Mn6 | (¾, ¾, ¾) | (1 1 1) |
| Ho7 | (0, -¼, -⅛) | (0 0 1) | Mn7 | (¼, ¼, ¾) | (-1 1 -1) |
| Ho8 | (0, ¼, ⅝) | (0 0 -1) | Mn8 | (¼, ¾, ¼) | (1 -1 -1) |
| Ho9 | (-⅛, 0, -¼) | (1 0 0) | | | |
| Ho10 | (⅝, 0, ¼) | (-1 0 0) | | | |
| Ho11 | (-¼, -⅛, 0) | (0 1 0) | | | |
| Ho12 | (¼, ⅝, 0) | (0 -1 0) | | | |

Figure S4 - Ordered magnetic moment vs transition temperature for $Ho_3Mn_xGa_{5-x}O_{12}$ for $x = 0.5$ and $x = 1$ respectively.

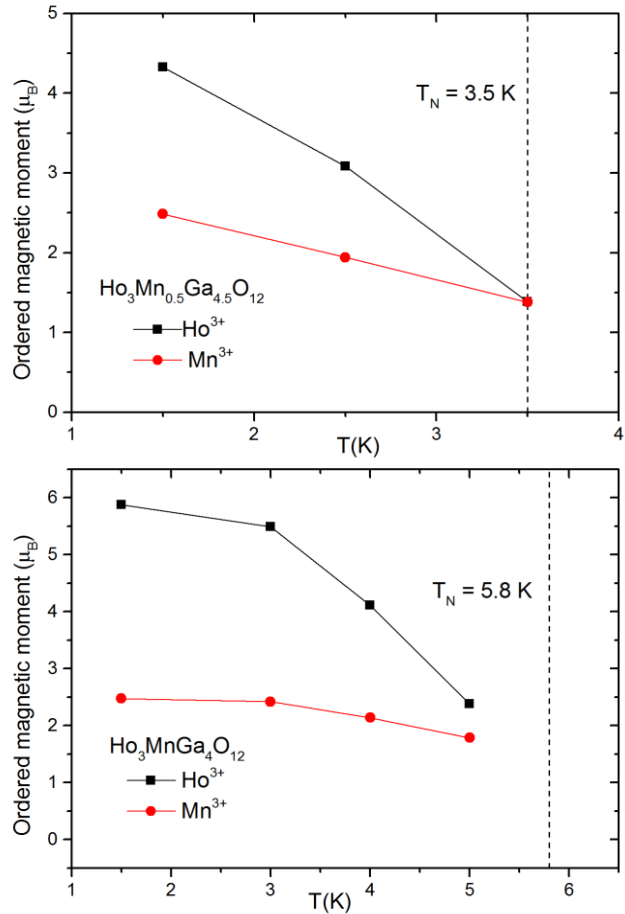